\documentclass[prl,aps,twocolumn,superscriptaddress]{revtex4-1}
\usepackage{graphicx}
\usepackage{subfigure}
\usepackage{bm}
\usepackage{amsmath}
\usepackage{amssymb}
\usepackage{tabularx}
\usepackage{wasysym}
\usepackage{color}
\usepackage{array}
\usepackage{multirow}
\usepackage{makecell}
\usepackage[version=3]{mhchem}
\usepackage{float}
\usepackage{color}
\usepackage{hyperref}
\usepackage{lipsum}
\usepackage{ulem}

\makeatletter
\newcommand{\colorcaption}[2][]{%
  \begingroup%
  \renewcommand{\@caption@fignum@sep}{ (color online). }%
  \caption[#1]{#2}%
  \endgroup%
  }
\makeatother

\makeatletter
\newcommand*{\rom}[1]{\expandafter\@slowromancap\romannumeral #1@}
\makeatother

\begin{document}
\title{A family tree of two-dimensional magnetic materials with tunable topological properties}

\author{Huisheng Zhang}
\affiliation{Key Laboratory of Magnetic Molecules and Magnetic Information Materials of the Ministry of Education and Research Institute of Materials Science, Shanxi Normal University, Linfen 041004, China}
\affiliation{International Center for Quantum Design of Functional Materials (ICQD), Hefei National Laboratory for Physical Sciences at Microscale, and CAS Center For Excellence in Quantum Information and Quantum Physics, University of Science and Technology of China, Hefei, Anhui 230026, China}
\author{Ping Cui}
\affiliation{International Center for Quantum Design of Functional Materials (ICQD), Hefei National Laboratory for Physical Sciences at Microscale, and CAS Center For Excellence in Quantum Information and Quantum Physics, University of Science and Technology of China, Hefei, Anhui 230026, China}
\affiliation{Key Laboratory of Strongly-Coupled Quantum Matter Physics, Chinese Academy of Sciences, School of Physical Sciences, University of Science and Technology of China, Hefei, Anhui 230026, China}
\author{Xiaohong Xu}
\thanks{Corresponding author: xuxh@sxnu.edu.cn}
\affiliation{Key Laboratory of Magnetic Molecules and Magnetic Information Materials of the Ministry of Education and Research Institute of Materials Science, Shanxi Normal University, Linfen 041004, China}
\author{Zhenyu Zhang}
\thanks{Corresponding author: zhangzy@ustc.edu.cn}
\affiliation{International Center for Quantum Design of Functional Materials (ICQD), Hefei National Laboratory for Physical Sciences at Microscale, and CAS Center For Excellence in Quantum Information and Quantum Physics, University of Science and Technology of China, Hefei, Anhui 230026, China}

\date{\today}

\begin{abstract}

Two-dimensional (2D) magnetic materials empowered with nontrivial band topology may lead to the emergence of exotic quantum states with significant application potentials. Here we predict a family tree of 2D magnetic materials with tunable topological properties, starting from the parental materials of CrI$_3$ and CrBr$_3$. The underlying design principle is that, by substituting the alternating sites of the Cr honeycomb lattice sandwiched between the halogen layers with V or Mn, the parental materials of trivial ferromagnetic insulators are ripe to be converted into topological systems. Specifically, our first-principles calculations show that, due to the elegant interplay between bandgap narrowing and spin-orbital coupling, CrI$_3$ branches into high-temperature quantum anomalous Hall insulators of CrVI$_6$ and CrMnI$_6$ with different topological invariants, while CrBr$_3$ branches into topological half-metals of CrVBr$_6$ and CrMnBr$_6$. Those novel 2D magnets are also shown to be easily exfoliated from their bulk counterparts. The present study is thus geared to advance the field of 2D magnetic materials into the topologically nontrivial realm.

\end{abstract}

\maketitle

\begin{figure}
\centering
\includegraphics[width=\columnwidth]{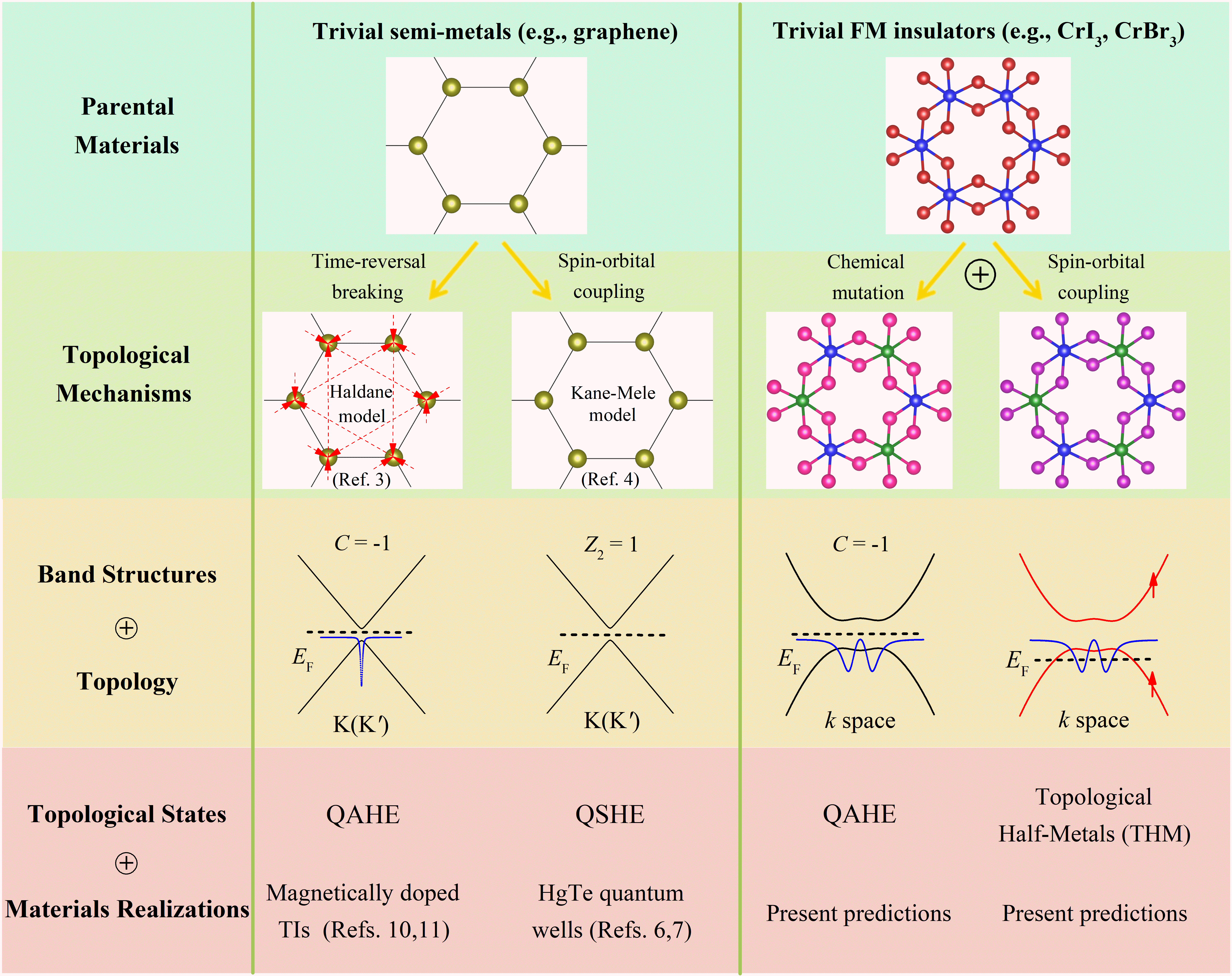}
\caption{(Left panels) Schematic diagrams of the Haldane and Kane-Mele models invoking respectively the time-reversal symmetry breaking and strong spin-orbital coupling in $sp$-electron based honeycomb lattices. Also shown are the corresponding band structures, topological quantum states, and materials realizations as discussed in the main text. (Right panels) A corresponding family tree on the emergence of topological states resulting from the interplay of bandgap narrowing and spin-orbital coupling upon chemical mutation of the trivial magnetic insulators of $d$-electron based honeycomb lattices. The strong predictions of the high-temperature, high-Chern number QAHE and novel topological half-metals are to be experimentally realized.}
\label{fig:figure1}
\end{figure}

Exploration of topologically nontrivial quantum states in two-dimensional (2D) or quasi-2D systems has attracted considerable research attention, in part because of their fundamental importance, and in part because such systems may have significant applications in future electronics and quantum computing. One celebrated example is the quantum Hall effect (QHE), which was first observed in a 2D electron gas confined within a quantum well under a high external magnetic field \cite{K. v. Klitzing}. Shortly after, the quantized Hall conductance was shown to be inherently connected to a topologically nontrivial state characterized by the TKNN invariant \cite{D. J. Thouless}. A significant conceptual advance was further made by Haldane, theorizing to achieve the QHE without the application of an external magnetic field, known as the QAHE \cite{F. D. M. Haldane}. Here, the role of the magnetic field is executed by the time-reversal symmetry breaking associated with the next-nearest-neighbor coupling of the electrons moving on a honeycomb lattice that in essence models closely the graphene lattice (left panel in Fig.~\ref{fig:figure1}). Starting from the same semi-metallic model of graphene, Kane and Mele took a different route by showing that the system could transit into another topological state of quantum spin Hall (QSH) insulator upon the inclusion of sufficiently strong spin-orbital coupling (SOC) \cite{C. Kane1} (left middle panel in Fig.~\ref{fig:figure1}). The corresponding topological invariant of the QSH state is $Z_2$ = 1 \cite{C. Kane2}.

Beyond the above conceptual advances \cite{F. D. M. Haldane,C. Kane1,C. Kane2}, the first experimental realizations of the QAHE and QSHE each took a distinctly different and remarkable materials pathway, as summarized in the bottom frame of Fig.~\ref{fig:figure1}. Specifically, a theoretical scheme to realize the QSHE was based on HgTe quantum wells with tunable well width \cite{B. A. Bernevig}, and the proposal was soon realized experimentally \cite{M. Konig}. In contrast, the QAHE was proposed much earlier, but its experimental realization was made only after the more recent discovery of 3D topological insulators (3DTIs) \cite{H. Zhang,Y. L. Chen}. Here, it was predicted that proper magnetic element doping will result in magnetic topological insulators such as Cr-doped Bi$_2$Se$_3$ thin films that could harbor the QAHE \cite{R. Yu}, and the prediction was confirmed experimentally upon proper generalization of the original recipe \cite{C. Z. Chang1}. Although some different schemes have been theoretically proposed \cite{M. Ezawa,J. Hu, Z. F. Wang, H. Pan, Z. Qiao, S. C. Wu, Y. Li, G. Xu, L. Dong, J. Liu, L. Si, P. Chen, H. Liu, C. Huang, Z. Liu, Q. Sun, Y. Hou, H. Fu}, the prevailing design principle for realizing the QAHE has been focused on magnetically doped 3DTIs, such as V-doped \cite{C. Z. Chang2}, V-I or Cr-V codoped \cite{S. Qi, Y. Ou}, and $\delta$ doped \cite{M. M. Otrokov1,Y. Gong,D. Zhang, J. Li, M. M. Otrokov2, Y. Deng} TIs.

 Despite several years of intensive efforts, the experimental temperatures to observe the QAHE remain exceptionally low: raised from 30 mK \cite{C. Z. Chang1} to $\sim$1 K \cite{ Y. Ou, Y. Deng}. To potentially overcome this standing hurtle, empowering nontrivial band topology to the recently discovered 2D ferromagnetic materials \cite{C. Gong, B. Huang} may offer an appealing new route to realizing high-temperature QAHE. Here, the 2D materials are intrinsically ferromagnetic at much higher transition temperatures, with time-reversal symmetry inherently broken. Furthermore, proper modification of such materials may render nontrivial band topology to these magnetic insulators, potentially resulting in topological magnetic insulators (namely, QAH insulators) or novel topological ferromagnetic metals or half-metals. Clearly, all such possible materials may have important applications in high performance spintronics devices.

In this study, we theorize a family tree of 2D magnetic materials with tunable topological properties, starting from the parental materials of CrI$_3$ and CrBr$_3$. Unlike the honeycomb lattice of graphene with explicit or implicit $sp$-electronic bands, here in the parental CrI$_3$ and CrBr$_3$, the honeycomb lattice of Cr sandwiched between the halogen layers is constructed with $d$-electronic bands. Detailed first-principles studies have already shown that such parental materials are trivial ferromagnetic insulators with significant band gaps ($\sim$0.89 eV for CrI$_3$ and $\sim$1.38 eV for CrBr$_3$) \cite{L. Webster}. To narrow the band gaps without compromising on the magnetic properties, we can replace alternating sites of the Cr honeycomb lattice by V or Mn, equivalent to hole or electron doping. Such severely bandgap narrowed or even closed systems should have much higher chances to be converted into the topologically nontrivial regime by the SOC \cite{G. Cao}, as unambiguously confirmed through first-principles calculations.

\begin{figure*}
\centering
\includegraphics[width=2\columnwidth]{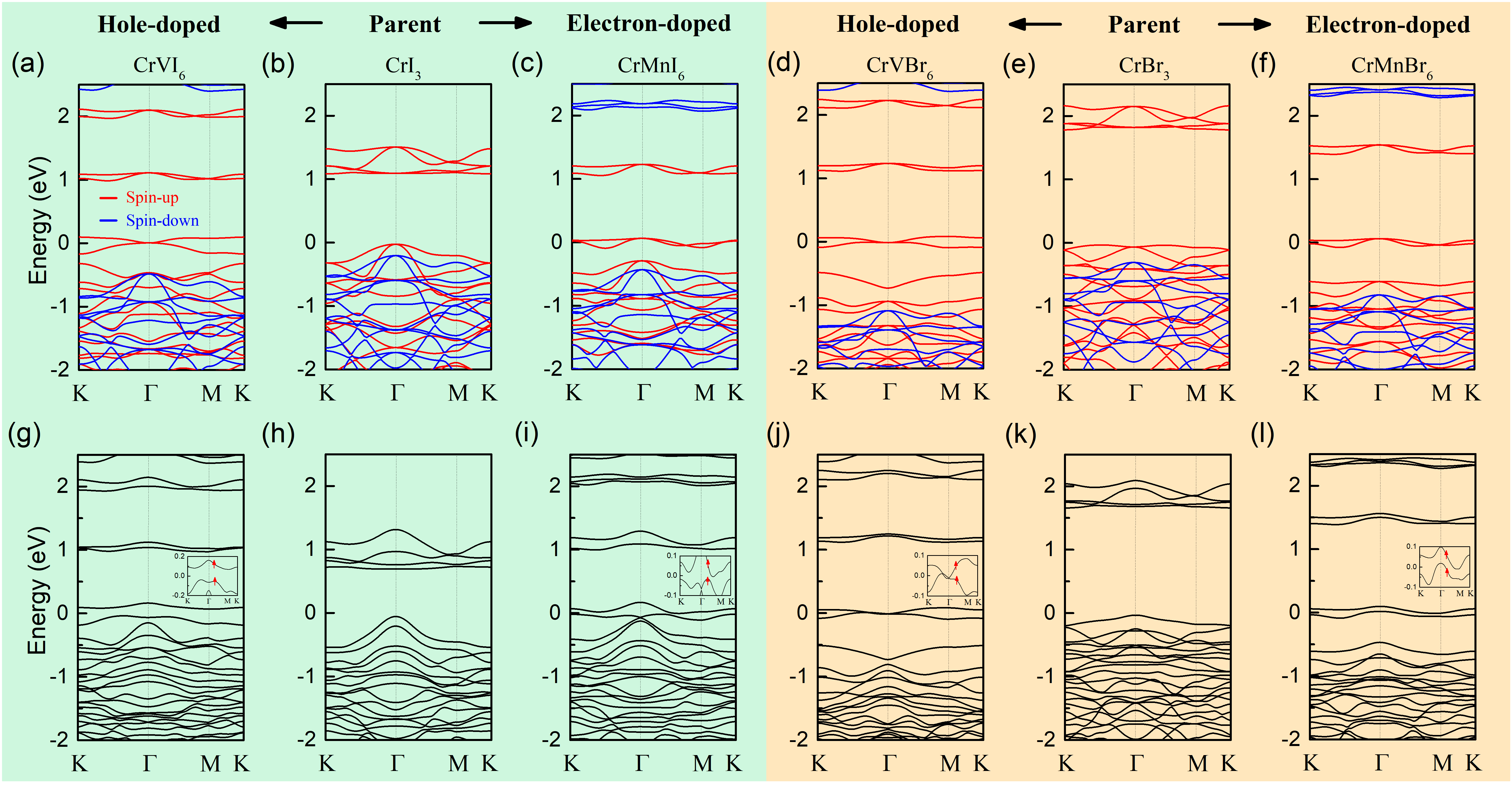}
\caption{Band structures of the (CrVI$_6$, CrI$_3$, CrMnI$_6$) and (CrVBr$_6$, CrBr$_3$, and CrMnBr$_6$) monolayers without (a-f) and with SOC (g-l). The red and blue lines in (a-f) represent the spin resolved bands. The insets in (g, i, j, l) illustrate the magnified bands near the Fermi levels, and the spin-polarized channels are indicated by the red arrows. The Fermi level is set at zero.}
\label{fig:figure2}
\end{figure*}

\begin{figure*}
\centering
\includegraphics[width=2\columnwidth]{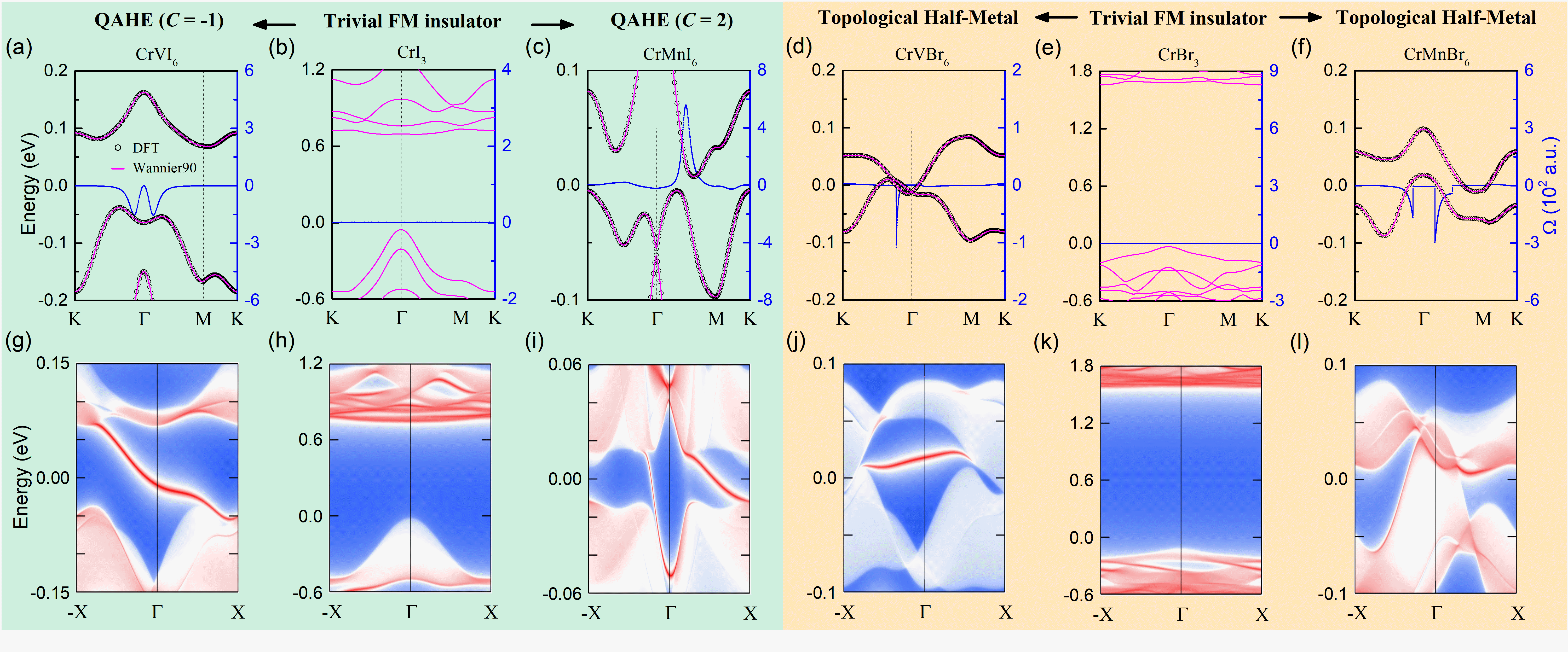}
\caption{Band structures of the (CrVI$_6$, CrI$_3$, CrMnI$_6$) and (CrVBr$_6$, CrBr$_3$, and CrMnBr$_6$) monolayers with SOC by using first-principles calculations and Wannier interpretations. The blue dots denote the Berry curvature (in atomic units (a.u.)). (g-l) Corresponding edge states in four of the six half-infinite monolayers.}
\label{fig:figure3}
\end{figure*}

We start with the parental materials of ultrathin CrI$_3$ and CrBr$_3$ films, which have been successfully fabricated from their bulk counterparts via mechanical exfoliation and shown to be ferromagnetic (FM) insulators with respective transition temperatures of $\sim$45 and 34 K in the monolayer regime \cite{B. Huang, Z. Zhang}. Our detailed calculations show that the pristine CrI$_3$ and CrBr$_3$ monolayers are trivial FM insulators, and their corresponding band structures and topological properties are shown in Figs.~\ref{fig:figure2} and ~\ref{fig:figure3}. Here, the Cr $d_{x^2-y^2,xy,z^2}$ orbitals are fully occupied, while the Cr $d_{xz,yz}$ orbitals are completely empty, resulting in the insulating states with sizeable band gaps. We emphasize that the Cr atoms form a honeycomb lattice (right top panel in Fig.~\ref{fig:figure1}) identical to that of graphene, suggesting that these trivial FM insulators are ripe to be converted into nontrivial FM materials upon proper manipulation.

Next we focus on how to introduce nontrivial band topology into the CrI$_3$ and CrBr$_3$ monolayers. When one of the two Cr atoms in a primitive cell is substituted by a neighboring transition metal element of Cr, an electron or hole is naturally introduced into these two monolayers (right middle panels in Fig.~\ref{fig:figure1}). In doing so, it is expected that some of the $d$-electronic bands could be moved to or created around the Fermi level, and the SOC effect can now be more effective in inducing nontrivial band topology in the V- or Mn-substituted CrI$_3$ or CrBr$_3$ monolayers. In particular, if a global band gap is still present for a specific combination, then it can host the QAHE.

Now, we explore the electronic structures of the CrVI$_6$, CrVBr$_6$, CrMnI$_6$, and CrMnBr$_6$ monolayers. For these four systems, the Cr and V (or Mn) atoms respectively occupy a triangular lattice, and the two lattices are further intertwined to form a new honeycomb lattice as depicted in the right middle panels of Fig. ~\ref{fig:figure1}. Figures ~\ref{fig:figure2}(a) and (c) give the band structures of CrVI$_6$ and CrMnI$_6$ monolayers respectively, exhibiting complete spin polarization around the Fermi level in the absence of SOC. Here, the non-isovalent substitutions and the preserved $C_3$ rotational symmetry collectively result in the half-metallic behavior. From the projected bands shown in Figs. S1 and S2 of the Supplementary Material \cite{Supplemental}, we observe that the new bands near the Fermi level are mainly contributed by the $d_{x^2-y^2,xy}$  orbitals of V and $d_{xz,yz}$ orbitals of Mn for CrVI$_6$ and CrMnI$_6$, respectively. When the SOC is switched on, both systems become insulators that may be intrinsic QAHE insulators, as confirmed by our detailed Chern number and edge states calculations presented later. The CrVI$_6$ monolayer hosts a global band gap of $\sim$0.1 eV, while a global band gap of $\sim$13 meV is obtained for CrMnI$_6$.

Similarly, the CrVBr$_6$ and CrMnBr$_6$ monolayers also host the half-metallic behavior without the SOC (Figs.~\ref{fig:figure2}(d) and (f)). In particular, their half-metallicity is preserved even upon inclusion of the SOC, because of the relatively weaker SOC of the Br atoms compared to the I atoms. The half-metallic bands near the Fermi levels are also contributed by the $d_{x^2-y^2,xy}$ orbitals of V and $d_{xz,yz}$ orbitals of Mn for CrVBr$_6$ and CrMnBr$_6$, respectively. These two systems are further characterized as topological half-metals, a striking and new prediction in the field of 2D materials. As a clear indication on the importance of the SOC, we have checked in Figs. S3 and S4 that, when the SOC strengths of Br are artificially enhanced towards that of I, band gaps can also be opened for the CrMnBr$_6$ and CrVBr$_6$ monolayers.

The detailed topological properties of CrVI$_6$, CrMnI$_6$, CrVBr$_6$, and CrMnBr$_6$ are determined by calculating the Berry curvatures and edge states of these systems (See Supporting Note 2 \cite{Supplemental}). Figures ~\ref{fig:figure3}(a-f) give the corresponding band structures of the six monolayers obtained from the Wannier90 package, which are nearly the same as that obtained from first-principles calculations. For the parental materials of CrI$_3$ and CrBr$_3$ (Figs.~\ref{fig:figure3}(b) and (e)), nearly zero Berry curvatures are present in the first Brillouin zone, and there is no edge state in the two half-infinite monolayers (Figs.~\ref{fig:figure3}(h) and (k). For CrVI$_6$ (Fig.~\ref{fig:figure3}(a)), negative Berry curvatures only appear around the $\Gamma$ point. By integrating these non-zero Berry curvatures, a Chern number of $C$ = -1 is obtained. In addition, we have further confirmed that for a half-infinite monolayer, a chiral state appears along the edge (Fig.~\ref{fig:figure3}(g)), demonstrating that the CrVI$_6$ monolayer is an intrinsic QAH insulator. The situation is richer for CrMnI$_6$. Here, positive and negative Berry curvatures coexist along the high symmetry paths (Fig.~\ref{fig:figure3}(c)). By integrating these non-zero Berry curvatures, a high Chern number of $C$ = 2 is obtained. Correspondingly, two chiral states appear along the edge of a half-infinite system, as displayed in Fig.~\ref{fig:figure3}(i). The detailed mechanism for the high Chern number of CrMnI$_6$ is also explored (See Supporting Note 3 \cite{Supplemental}). Furthermore, the CrVBr$_6$ and CrMnBr$_6$ monolayers are characterized as topological half-metals (Figs.~\ref{fig:figure2}(j) and (l)), and a chiral state appears along the edge of each of them (Figs.~\ref{fig:figure3}(j) and (l)). The projected bands of the CrVI$_6$, CrMnI$_6$, CrVBr$_6$, and CrMnBr$_6$ monolayers with SOC are displayed in Fig. S6.

The magnetic properties of the six monolayers are systematically investigated next. Our calculations reveal that the total energy of the FM structure is consistently lower than that of the antiferromagnetic (AFM) structure for each of the six monolayers, as listed in Table I. We have also calculated the magnetic anisotropy energy (MAE) of the six monolayers defined by MAE = $E_{\parallel}-E_{\bot}$, with $E_{\parallel}$ and $E_{\bot}$ denoting the total energies associated with the magnetic moment parallel and perpendicular to the plane of the 2D materials, respectively. As shown in Table I, the obtained MAE is positive for all the six systems, namely, the easy magnetization axis is out of plane, as expected for their FM nature beyond the Mermin-Wagner theorem \cite{N. D. Mermin}. Quantitatively, the magnetization of the CrMnI$_6$ and CrMnBr$_6$ monolayers is $\sim$7.0 $\mu_B$ per unit cell, larger than that of CrI$_3$ and CrBr$_3$ (with 6.0 $\mu_B$ per unit cell), as a Cr$^{3+}$ or Mn$^{3+}$ ion possesses three or four unpaired electrons, respectively. Similarly, the magnetization of CrVI$_6$ and CrVBr$_6$ is $\sim$5.0 $\mu_B$ per unit cell. These findings are insensitive to whether the SOC effect is included or not.

\begin{table*}
  \centering
  \caption{Structural, electronic, magnetic, and topological properties of the six monolayers. Here, $a$, $M_{tot}$, and MAE represent the optimized lattice constant, total magnetic moment, and magnetic anisotropy energy of a monolayer, respectively. $\Delta_{AFM-FM}$ denotes the energy difference between the AFM and FM configurations.}
  \begin{ruledtabular}
   \begin{tabular}{cccccccc}
      & CrVI$_6$ & CrI$_3$ & CrMnI$_6$ & CrVBr$_6$ & CrBr$_3$ & CrMnBr$_6$ \\
     \hline
     $a$(\AA) & 7.02 & 7.00 & 7.06 & 6.49 & 6.44 & 6.50 \\
     $\Delta_{AFM-FM}$ (meV) & 43.9 & 54.8 & 41.2 & 35.7 & 45.6 & 36.1 \\
     $M_{tot}$ ($\mu_B$) & 5.0 & 6.0 & 7.0 & 5.0 & 6.0 & 7.0 \\
     MAE (meV) & 0.25 & 0.28 & -0.36 & 0.28 & 0.21 & 0.32 \\
     $T_c$ (K) & 75 & 54 (45\footnotemark[1]) & 87 & 63 & 51 (34\footnotemark[2]) & 59 \\
     \makecell{Topological\\states} & QAHE & \makecell{Trivial\\insulator} & QAHE & \makecell{Topological\\Half-Metal} & \makecell{Trivial\\insulator} & \makecell{Topological\\Half-Metal} \\
     \makecell{Topological\\invariant} & $C$ = -1 & $C$ = 0 & $C$ = 2 & /\footnotemark[3]) & $C$ = 0 & / \\
        \end{tabular}
   \end{ruledtabular}
   \footnotetext[1]{The experimental $T_c$ in Ref.\cite{B. Huang}.}
   \footnotetext[2]{The experimental $T_c$ in Ref.\cite{Z. Zhang}.}
   \footnotetext[3]{$"/"$ denotes that the system hosts metallic behavior, which cannot be identified by a Chern number.}
\end{table*}

For such 2D FM materials, the transition temperature $T_c$ is another important measure. Here we use Monte Carlo simulations to estimate $T_c$ of the CrVI$_6$, CrMnBr$_6$, CrVBr$_6$, and CrMnBr$_6$ monolayers (see Supporting Note 4), as shown in Table I. For comparison, we have also estimated the $T_c$ of the CrI$_3$ and CrBr$_3$ monolayers to be $\sim$54 and 51 K, which are somewhat higher than that of the experimental results ($\sim$45 \cite{B. Huang} and 34 K \cite{ Z. Zhang}). As the nontrivial band gaps of CrVI$_6$ and CrMnI$_6$ are much larger than the energy scale of their corresponding FM orders, the latter determines the observation temperature of the QAHE, demonstrating that these two monolayered systems are QAH insulators at much higher temperatures than that of Cr-V codoped and $\delta$ doped TIs \cite{Y. Ou, Y. Deng}. In particular, the CrMnI$_6$ monolayer is a high-temperature QAHE system with a high Chern number of $C$ = 2. The detailed electronic, magnetic, and topological properties of the six monolayers are compared in Table I.

For potential experimental realization of the four monolayers, we have confirmed that they are dynamically and thermodynamically stable (See Supporting Note 6 \cite{Supplemental}). To further assess the feasibility of fabricating these four monolayers, we have also examined their bulk counterparts, and found that their interlayer distances are $\sim$3.10 \AA, signifying the van der Waals (vdW) nature of the interlayer coupling, and the exfoliation energies of those monolayers are comparable to that of graphene (See Supporting Note 7 \cite{Supplemental}). Furthermore, we have investigated the electronic and magnetic properties of these four bulk systems as shown in Fig. S10, finding that FM order is the ground state for each bulk system, similar to the parental materials of CrI$_3$ and CrBr$_3$. A key difference is that, while the interlayer coupling is AFM in bilayer CrI$_3$ \cite{B. Huang} or CrBr$_3$ \cite{Z. Zhang}, FM ordering is the ground state in each of the bilayers of CrVI$_6$, CrMnI$_6$, CrVBr$_6$, and CrMnBr$_6$. This feature naturally renders CrVI$_6$ and CrMnI$_6$ another merit as platforms for observing the QAHE. In contrast, the interlayer coupling of two nearest septuple layers of MnBi$_2$Te$_4$ \cite{M. M. Otrokov1,D. Zhang, J. Li, M. M. Otrokov1} is AFM, demanding that the QAHE can only be realized in odd-layered MnBi$_2$Te$_4$ \cite{Y. Deng}.

As there are two types of magnetic elements in the CrVI$_6$, CrVBr$_6$, CrMnI$_6$, or CrMnBr$_6$ monolayer, some disordered distributions of Cr and V (or Mn) are likely to be introduced during fabrication, potentially causing degradation of the magnetic  and topological properties. To address this vital practical issue, we have investigated the electronic, magnetic, and topological properties of the disordered four monolayers in which a V (or Mn) and Cr atom pair exchange their locations in the honeycomb lattice as shown in Fig. S11, corresponding to a 25\% disorder. We found that the total energies of the ordered configurations are always lower than that of the disordered configurations. More importantly, our calculations show that the FM states are well preserved in the disordered configurations (see Fig. S11). Taking the CrVI$_6$ monolayer as a specific example, its topological state is also well preserved, as shown in Fig. S12. More intriguingly, here the disorder can even induce a phase transition from the topological half-metal to a QAH insulator, namely, a global band gap of $\sim$15 meV is obtained in a disordered CrVBr$_6$ monolayer (see Fig. S12). These cross checks ensure that the topological state in the CrVI$_6$ monolayer is robust against disorder and physically feasible to be realized.

In summary, we have identified a family tree of 2D magnetic materials with tunable topological properties, starting from the parental materials of CrI$_3$ and CrBr$_3$ through proper chemical mutations. In particular, our first-principles calculations show that, due to the delicate interplay between bandgap narrowing and spin-orbital coupling, CrI$_3$ branches into high-temperature quantum anomalous Hall insulators of CrVI$_6$ and CrMnI$_6$, while CrBr$_3$ branches into topological half-metals of CrVBr$_6$ and CrMnBr$_6$. We have also shown that those novel 2D magnets can be readily exfoliated from their bulk counterparts, as supported by preliminary experiments. The present study is expected to drastically advance the field of topological 2D magnetic materials.

This work was supported by the National Key R\&D Program of China (Grant Nos. 2017YFB0405703 and 2017YFA0303500) and the National Natural Science Foundation of China (Grant Nos. 11804210, 61434002, 51871137, 11634011, 11722435, and 11974323), the Strategic Priority Research Program of Chinese Academy of Sciences (Grant No. XDB30000000), and the Anhui Initiative in Quantum Information Technologies (Grant No. AHY170000).

\end{document}